\documentclass{sig-alternate}

\usepackage{epsfig,subcaption,endnotes,svg, textcomp, float, algorithm, algpseudocode, url, balance}
\usepackage{paralist}
\usepackage{multirow}
\usepackage{dcolumn}
\usepackage{nameref}
\usepackage{color}
\usepackage{microtype}
\usepackage{pifont}

\newcommand{\avantguard}{\textsc{Avant-Guard}}

\numberofauthors{4}

\author{
	Moreno Ambrosin, Mauro Conti\thanks{\scriptsize Mauro Conti is supported by a Marie Curie Fellowship funded by the European Commission under the agreement No. PCIG11-GA-2012-321980. This work is also partially supported by the TENACE PRIN Project 20103P34XC funded by the Italian MIUR, and by the Project ``Tackling Mobile Malware with Innovative Machine Learning Techniques'' funded by the University of Padua.}, Fabio De Gaspari,\\
	\affaddr{University of Padua, Italy}\\
	{\{surname\}@math.unipd.it}\\
	{fabio.degaspari@studenti.unipd.it}
	\and
	Radha Poovendran\\
	\affaddr{University of Washington, USA}\\
	{rp3@uw.edu}	
}

\newfont{\mycrnotice}{ptmr8t at 7pt}
\newfont{\myconfname}{ptmri8t at 7pt}
\let\crnotice\mycrnotice%
\let\confname\myconfname%

\permission{Permission to make digital or hard copies of all or part of this work for personal or classroom use is granted without fee provided that copies are not made or distributed for profit or commercial advantage and that copies bear this notice and the full citation on the first page. Copyrights for components of this work owned by others than ACM must be honored. Abstracting with credit is permitted. To copy otherwise, or republish, to post on servers or to redistribute to lists, requires prior specific permission and/or a fee. Request permissions from permissions@acm.org.}
\conferenceinfo{ASIA CCS'15,}{April 14--17, 2015, Singapore.}
\copyrightetc{Copyright \copyright~2015 ACM \the\acmcopyr}
\crdata{978-1-4503-3245-3/15/04\ ...\$15.00.\\
http://dx.doi.org/10.1145/2714576.2714612}

\begin{document}

\title{LineSwitch: Efficiently Managing Switch Flow \\in Software-Defined Networking while \\Effectively Tackling DoS Attacks}

\maketitle

\begin{abstract}

Software Defined Networking (SDN) is a new networking architecture which aims to provide better decoupling between network control (control plane) 
and data forwarding functionalities (data plane). 
This separation introduces several benefits, such as a directly programmable and (virtually) centralized network control. 
However, researchers showed that the required communication channel between the control and data plane of SDN 
creates a potential bottleneck in the system, introducing new vulnerabilities.
Indeed, this behavior could be exploited to mount powerful attacks,
such as the {\em control plane saturation attack}, that can severely hinder the performance of the whole network.

In this paper we present LineSwitch, an efficient and effective solution against control plane saturation attack. 
LineSwitch combines SYN proxy techniques and probabilistic blacklisting of network traffic. 
We implemented LineSwitch as an extension of OpenFlow, the current reference implementation of SDN, 
and evaluate our solution considering different traffic scenarios (with and without attack).
The results of our preliminary experiments confirm that, compared to the state-of-the-art, LineSwitch reduces the time overhead up to 30\%, while ensuring the same level of protection.

\end{abstract}



\section{Introduction}
\sloppy
The great increase in demand for flexibility and automation in key Information Technology sectors is leading to the rise of new paradigms that greatly simplify the management of network infrastructures. 
One of such paradigms is Software Defined Networking (SDN). Unlike the current network infrastructure, SDN decouples the network layer (i.e., the {\em control plane}) and the data layer (i.e., the {\em data plane}) functionalities in separate entities. This separation allows a (virtually) centralized and directly programmable network control, while at the same time reduces and abstracts the complexity of network devices. 

The most widely adopted instantiation of the SDN paradigm is OpenFlow (OF)~\cite{McKeown:2008:OEI:1355734.1355746,openflow_whitepaper}. 
OF provides a standard communication interface between the data plane, and the control plane, 
and introduces the concept of {\em flows} to identify the network traffic~\cite{openflow_whitepaper}. 
Each OpenFlow entity, namely {\em OF switch}, maintains a set of {\em flow tables}. They specify the rules that the 
OF switch uses to perform the routing of packets. 
Moreover, flow tables are organized in a pipeline, which is traversed by the switch every time a packet is received. 
The control plane can program the flow tables of the OF switches either statically or dynamically.
In the latter case, if an OF switch does not have a matching rule in its pipeline for a new incoming packet, 
it must contact the control plane to retrieve a new rule~\cite{openflow_switch_specs}. 
Unfortunately, while enabling a flexible network management, the required extensive communication between control and data plane might result in poor scalability. Moreover, it introduces a serious vulnerability that can be exploited to overload the control plane with flow requests: the resulting attack is called {\em control plane saturation}~\cite{AvantGuard,of-guard}, and can be easily performed, for example, through SYN flooding~\cite{Peng:2007:SND:1216370.1216373}.
By overloading the control plane, this attack incapacitates the target OF switch, which will not be able to retrieve rules for new network flows. Furthermore, if the controller manages more than a single switch, the attacker might hinder an even larger part of the network~\cite{AvantGuard,Kloti,of-guard,Benton:2013:OVA:2491185.2491222}.

Recently, Shin et al.~\cite{AvantGuard} proposed \avantguard, a countermeasure against the control plane saturation attack.
\avantguard~introduces a new module into the OF switch, called {\em connection migration} module, 
which protects the switch and the controller from saturation attacks performed by SYN flooding, while at the same time being transparent to the end hosts.
With this module, each OF switch acts as a SYN proxy during the TCP handshake stage of a connection, effectively shielding the controller from possible floods. 
Unfortunately, while being beneficial in the general case, this technique introduces new subtle vulnerabilities and heavy limitations. 
Indeed, we will show that when running \avantguard, the state the OF switch needs to maintain in order to proxy each connection can lead to a new Denial of Service attack, that we refer to as {\em buffer saturation attack}. Furthermore, the transparency required with respect to the end hosts limits the number of connections that the switch can proxy to the number of available TCP port numbers.

\paragraph*{Our Contribution} In this paper, we make the following contributions:
\begin{compactitem}

	\item We identify and discuss some unintended vulnerabilities of one of the recently 
	proposed schemes against the control plane saturation attack that, for the best of our knowledge, 
	represents the state-of-the-art solution against this threat.
	
	\item We propose a novel attack, which we name {\em buffer saturation attack}.
	Our attack exploits some of the identified vulnerabilities introduced by the state-of-the-art solution
	for the control plane saturation attack. 
	As confirmed by our analysis, {\em buffer saturation attack} is both realistic and simple to run, 
	and leads to significant network performance degradation.
	
	\item We propose LineSwitch, 
	a new efficient and effective solution to mitigate the control 
	plane saturation attack. LineSwitch greatly reduces the effects of this 
	attack, while at same time protects the network from the {\em buffer saturation attack}.
	
	\item We did a preliminary evaluation of our solution, which confirms the effectiveness of LineSwitch against
	the {\em control plane saturation attack}. Moreover, our experiments show a significant reduction of the time 
	overhead (up to 30\%) when compared to the state-of-the-art.

\end{compactitem}

\paragraph*{Organization}
The remaining of this paper is organized as follows. In Section~\ref{sec:background} we provide some background knowledge on SDN, and on the SYN flooding attack and its possible countermeasures. 
In Section~\ref{sec:related_work} we revise some related work in the area of DoS attacks and defense in Software Defined Networking. 
Moreover, we provide a brief introduction of \avantguard, which represents the state-of-the-art solution against the control plane saturation attack. 
In Section~\ref{sec:connection_migration_analysis} we analyze the limitations of the current state-of-the-art, 
and introduce a new possible attack, i.e., the buffer saturation attack. 
Section~\ref{sec:lineswitch} describes LineSwitch, our countermeasure against the control plane saturation attack in SDN, 
while in Section~\ref{sec:evaluation} we provide a preliminary evaluation of its effectiveness. 
Finally, in Section~\ref{sec:conclusion} we draw our conclusions.

\section{Preliminaries}\label{sec:background}

In this section we introduce some concepts that will be used in the remaining of the paper: Software Defined Networking (Section~\ref{sec:sdn_preliminaries}) and the SYN flooding attack (Section~\ref{sec:syn_flooding}).

\subsection{Software Defined Networking (SDN)}\label{sec:sdn_preliminaries}
\begin{figure}[ht]
	    \centering
	    \includegraphics[width=0.65\columnwidth]{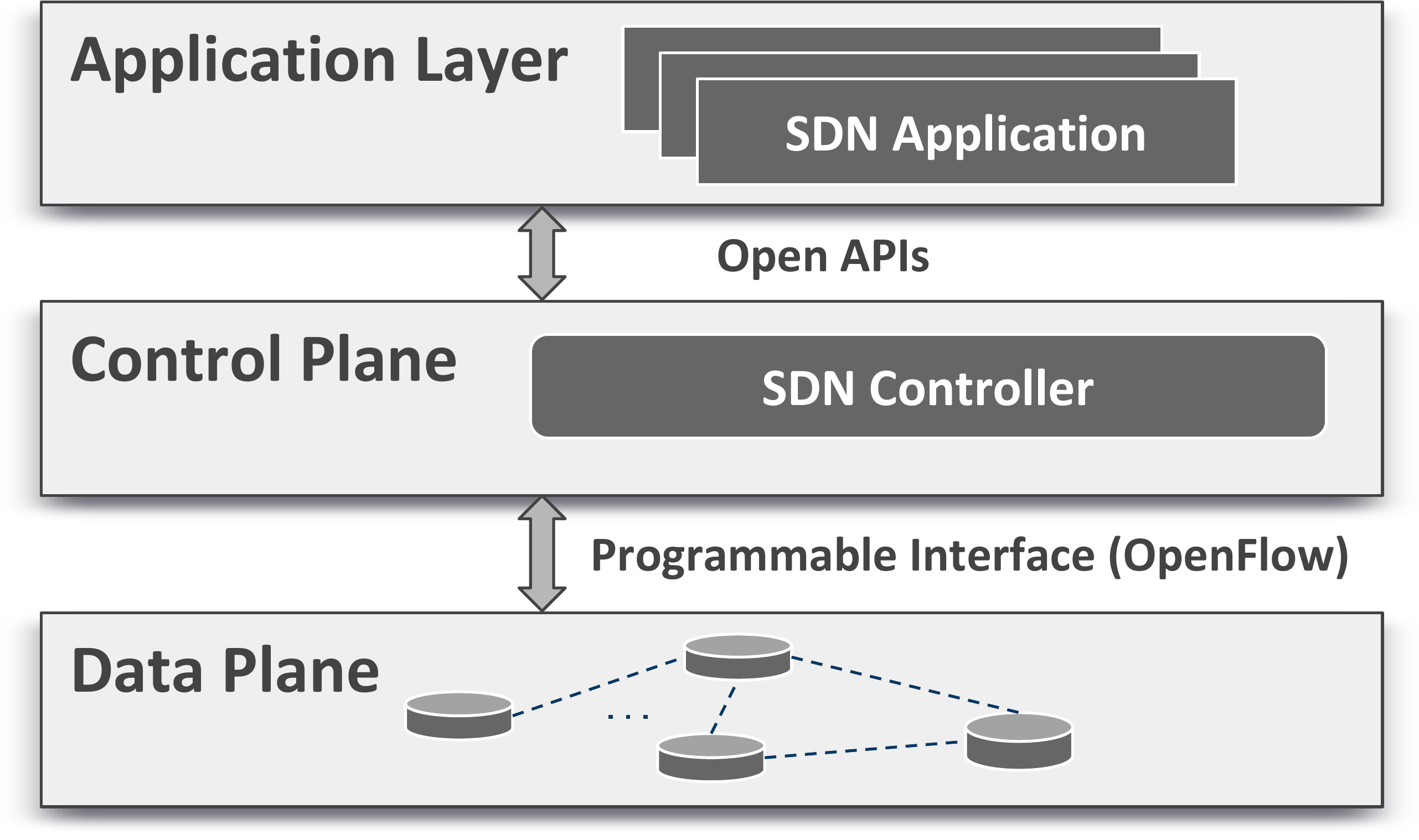}
	    \caption{SDN Architecture.}
	    \label{fig:SDN}
	\end{figure}

Software Defined Networking (SDN) has emerged as a new network paradigm aimed at providing higher flexibility in network research, development and operation. The core concept behind SDN is the separation of two distinct aspects of networking that in today's architecture are blended together: the {\em network control} and the {\em forwarding functions}.
The SDN architecture postulates that these two logically separated aspects of networking are decoupled in two corresponding layers, respectively the {\em control plane} and the {\em data plane}.
Figure~\ref{fig:SDN} provides a high-level representation of the SDN architecture.

The control plane provides a directly programmable layer that acts as an interface to the data plane for applications, abstracting both the complexity of the underlying network infrastructure and the communication between the two planes.
Using the control plane as a middleware, it is possible to easily deploy a great variety of network management applications, that act independently from the physical network devices. 
Moreover, the control plane offers a logical centralized system that controls and accesses data from all network devices, effectively offering a global view of the network infrastructure.
OpenFlow (OF)~\cite{openflow_whitepaper} is the reference implementation of the SDN paradigm. It defines a standard communication interface between the control plane and the data plane. 
With OpenFlow, the routing is performed based on traffic {\em flows}; each OF network entity (i.e., {\em OF switch}) maintains one or more {\em flow tables}, that are used to route incoming packets, and an OF channel to an external controller. 
The control plane can program the physical devices through a series of {\em flow rules}~\cite{openflow_switch_specs}, 
that are installed inside the flow tables. Such rules specify which actions a switch will perform on a specific network flow.
For each unique network flow, or group of flows, there will be a corresponding flow table entry (i.e., a flow rule). 
Once an OF switch receives a packet, it matches the packet header against the pipeline of its local flow tables, which will dictate what actions will be applied to that specific flow.
The control plane populates the flow tables of each OF switch either statically, by pre-installing a set of flow rules into the flow tables, 
or dynamically, by allowing a switch-driven run-time installation of new flow rules~\cite{openflow_whitepaper}. 
In the latter case, if an OF switch does not have any matching rule for a new incoming flow, it forwards the corresponding packet header to the controller, 
which at this point can install a new rule for that flow into the switch~\cite{openflow_switch_specs}.
Through the combination of different flow rules, a controller can define a broad range of actions, 
from the standard routing of a packet to a more complex analysis, involving forwarding the packet to the controller~\cite{openflow_switch_specs}.


\subsection{SYN Flooding Attack}\label{sec:syn_flooding}
SYN flooding attacks are one of the most widespread and effective DoS attacks~\cite{Peng:2007:SND:1216370.1216373}. The effectiveness of a SYN flooding attack is based on the state that TCP stacks allocate when a connection request (i.e., a SYN packet) is received. 
Such state, called {\em Transmission Control Block} (TCB)~\cite{RFC793}, is retained in memory for a certain amount of time, even if the TCP handshake is not completed by the client. This is needed to handle the case in which the ACK packet, needed to complete the connection, is lost due to network congestion. 
Once a critical amount of half-open connections is reached, the available memory reserved to the TCBs is saturated, and no new connections can be established.
In general, in order to make it difficult for the victim to detect the attack and its source, an attacker performs a SYN flooding attack using spoofed IP addresses. In this way, the victim will not be able to identify the source of the attack, and therefore, it will not be able to automatically stop the SYN flooding. 
Indeed, new incoming SYN packets could belong to legitimate connections, which the victim can not distinguish from the malicious traffic.
Researchers and industry proposed several countermeasures against the SYN flooding attack~\cite{RFC4987,Peng:2007:SND:1216370.1216373}. However, none of them offers a definitive solution to this important problem. 

When applied to SDN, the SYN flooding attack does not aim at saturating internal data structures. 
Rather, the goal is to exploit data to control plane communication to saturate the controller by generating a huge number of new network flows. 
For each flow, the OF switch will have to contact the controller, which will need to analyze the flow information, prepare and then send an OF response to the switch. 
If the rate of the flood is high enough, the controller will not be able to keep up with the attack and will be incapacitated from serving legitimate network flows.


\section{Related Work}\label{sec:related_work}

In this section, we provide a brief overview of the main research studies related to Denial of Service (DoS) attacks in SDN. 
Due to space limitations, in this section we focus only on DoS attacks against the control plane of an SDN network.
In~\cite{Peng:2007:SND:1216370.1216373}, Peng et al. provided a first feasibility study for Denial of Service attacks on SDN control plane. In~\cite{Kreutz:2013:TSD:2491185.2491199}, Kreutz et al. analyzed the SDN architecture, identifying critical aspects and possible new attack vectors, while Kloti et al.~\cite{Kloti} assessed some vulnerabilities that affects OpenFlow~\cite{openflow_whitepaper}, i.e., the possibility for an attacker to mount DoS attacks on the control plane, and to disclose potentially sensitive information by using timing analysis techniques.
In~\cite{AvantGuard}, Shin et al. propose \avantguard, a solution that addresses architectural flows of the original OpenFlow protocol by altering the flow management at the data plane level~\cite{of-guard, AvantGuard}. 
In particular, the authors focused on solving OpenFlow's control plane saturation attack vulnerability. 
To the best of our knowledge, \avantguard~\cite{AvantGuard} represents the state-of-the-art for tackling the control plane saturation attack in SDN. 
For this reason, in the remaining of this section we briefly describe this solution. 
We will further compare \avantguard~against our solution when evaluating our proposal (see Section~\ref{sec:evaluation}).

\avantguard~\cite{AvantGuard} adds two extensions to the standard OpenFlow protocol: 
(1) a {\em Connection Migration} module, which limits the effect of the control plane saturation attack based on SYN flooding by proxying the incoming TCP requests, and 
(2) the {\em Actuating Trigger} module, which allows the controller to limit the number of control messages required to collect network statistics.
Since the focus of this paper is on solving the control plane saturation attack, in the remaining of this paper we will focus only on the connection migration module of \avantguard, that we briefly describe in this section. Moreover, in Section~\ref{sec:connection_migration_analysis} 
we will provide a security analysis of the connection migration module.

As introduced in Section~\ref{sec:sdn_preliminaries}, whenever an OF switch receives an inbound network flow for which it has no flow rule, it will forward the packet to the control plane. This behavior holds even for SYN packets: indeed, for each received SYN packet not matching a flow rule, an OF switch will contact the controller to obtain a corresponding rule.
The proposed connection migration module of \avantguard~addresses this problem at the data plane level, 
by having the OF switch act as a SYN proxy.

\begin{figure}[!ht]
	    \centering
	    \includegraphics[width=0.83\linewidth]{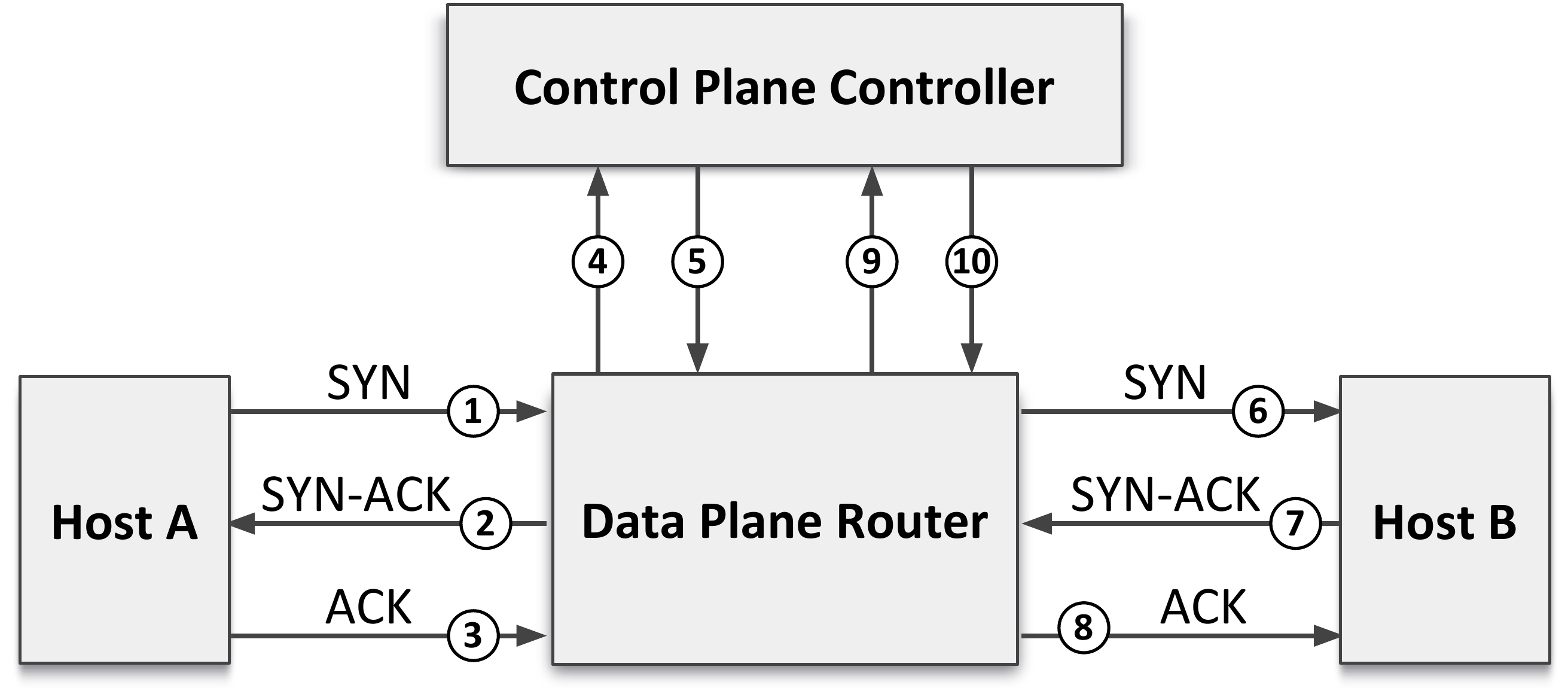}
	    \caption{\avantguard~\cite{AvantGuard} -- Connection Migration.}
	    \label{fig:conn_migration}
	\end{figure}

This process is articulated in four phases (see Figure~\ref{fig:conn_migration}):

		{\em Classification phase.} When an OF switch receives a SYN packet belonging to an unknown flow (action (1) in Figure~\ref{fig:conn_migration}), instead of forwarding it to the control plane, the switch acts as a proxy, engaging the client in a stateless TCP handshake through SYN Cookies (actions (2) and (3) in Figure~\ref{fig:conn_migration}).
		
		{\em Report phase.} If the client completes the TCP handshake, the switch then forwards the new flow to the controller (action (4) in Figure~\ref{fig:conn_migration}) and waits for a new rule that defines how it should be handled (action (5) in Figure~\ref{fig:conn_migration}).
		
		{\em Migration phase.} If the controller allows the migration, the switch initiates a TCP handshake with the destination host (actions (6), (7) and (8) in Figure~\ref{fig:conn_migration}). The OF switch further reports the result of the handshake to the control plane (actions (9) and (10) in Figure~\ref{fig:conn_migration}).
		
		{\em Relay phase.} If the handshake is successful, the switch forwards all the subsequent messages between the client and the destination host.

The foremost advantage of connection migration, is the classification mechanism. 
Only complete TCP flows will be reported to the control plane, effectively shielding it from SYN 
flooding attacks performed with spoofed IP addresses, and greatly mitigating the threat of link saturation. 
For non-spoofed TCP flows, the result is that any $\left\{ {IP, port } \right\}$ combination will appear to be valid, 
effectively converting the network to a whitehole network and preventing an attacker from mapping possible targets.
Moreover, the consequent SYN flooding vulnerability at data plane level is addressed by the use of SYN Cookies~\cite{syncookies}. 
Since the SYN Cookie algorithm does not need to maintain state for connection requests, 
there is no need for storing information in the OF switch for failed TCP connections.

Generally, it is possible to infer the use of connection migration by analyzing the round trip time of a SYN packet. 
An adversary might use this information to flood the data plane with complete TCP handshakes, 
forcing the switch to forward each of them to the control plane.
To solve this issue, \avantguard~provides a modification to the basic connection migration module, 
namely {\em delayed connection migration}~\cite{AvantGuard}, which requires the initiator of the 
communication to send the first valid packet, before forwarding the packer header to the control plane.


\section{Limitations of the Connection Migration module}\label{sec:connection_migration_analysis}

The connection migration module of \avantguard~\cite{AvantGuard} is indeed a valid solution against the control plane saturation attack. 
Moreover it also shields end hosts from SYN flooding attacks and, by replying unconditionally to every received SYN packet, it prevents port scanning attacks.
Unfortunately, along with the above desirable properties, the connection migration module of \avantguard~introduces new vulnerabilities too. 
In particular, we identified two distinct vulnerabilities that can lead to a shutdown of the OF switch.
First, the state that the switch must keep in order to implement the connection migration module of \avantguard~can be exploited by an attacker, which can try to fill the allotted buffers and incapacitate the OF switch (see Section \ref{sec:proxying_needs_state}). Second, the use of SYN proxy limits the number of connections that can be forwarded. In particular, the maximum number of connections forwarded to a specific $\left\{ {IP, port } \right\}$ pair is $64513$ (see Section \ref{sec:proxy_ports}).

In what follows we provide an in-depth analysis, as well as the scheme for possible attacks, for each of the above points.

\subsection{Proxying Requires State}\label{sec:proxying_needs_state}

A switch implementing the connection migration module of \avantguard~needs to maintain some state, for the whole duration of the TCP connection.
In particular, acting as a proxy between two communicating hosts, $A$ and $B$, a switch $R$ should execute the following three operations: 

		(1) Once received a SYN packet from host $A$, with Initial Sequence Number~\cite{RFC793} $ISN_A$, $R$ will respond with a SYN-ACK packet with a spoofed address, i.e., using host $B$ address. The ACK sequence number will be $ISN_A+1$ and the sequence number will be a random number $ISN_R$. Note that, according to TCP protocol specifications~\cite{RFC6528, Morris1985} each ISN is computed in a non-predictable way. Therefore, it is impossible for $R$ to predict the ISN that $B$ would generate and, as a consequence $ISN_R$ will be different from the sequence number $B$ will use.
		
		(2) Upon receiving the permission to migrate the connection, switch $R$ will start a handshake with host $B$ by sending a SYN packet with sequence number $ISN_A$. Note that at this stage, $R$ must use its IP address to establish the TCP session since it has no guarantees that the reply from host $B$ will follow the same path through switch $R$ on the way back. 
		
		(3) Once the SYN-ACK packet from host B (with an ACK number $ISN_A+1$ and a sequence number $ISN_B$) is received, switch R finalizes the connection sending an ACK with number $ISN_B+1$ to B.
		
To maintain the connection migration transparent to the end hosts, $R$ must perform a {\em sequence number translation} (and analogously an ACK translation), for the packets exchanged by $A$ and B. Moreover, for each host connecting to the same $\left\{ {IP, port} \right\}$ pair, the switch needs to use a distinct port number to migrate the connection in order to later match the response packets to the correct host on the way back.
Consequently, for each connection the switch needs to store the following information:

\begin{center}
	$\left\{ {IP_{src}, port_{src}, port_R, \delta_{seq}} \right\}$,
\end{center}

where $IP_{src}$ and $port_{src}$ are respectively source address and port of the initiator of the connection, 
$port_R$ is the port number used by the router in the migration and $\delta_{seq}$ is the difference between the ISN used by the router and the ISN used by the destination host.

The translation table required by the switch to act as a proxy gives an attacker an easy mean for mounting a {\em buffer saturation attack}.
Indeed, the attacker simply needs to open several complete TCP connections through the target OF switch to a given host.
Note that each of these connections will need state to be stored on the switch for translation. 
Therefore, if the number of connections is large enough, the portion of memory dedicated to that data structure will be saturated, 
incapacitating the switch from serving any further valid connection.

\subsection{Limit on the Number of Connections}
\label{sec:proxy_ports}
As we already stated in Section~\ref{sec:proxying_needs_state}, when a connection from hosts $A$ to $B$ is proxied by switch $R$, 
all the packets translated by the switch to the destination $B$ will have the IP address of $R$ and a port number which will be different 
from the original one used by A.
This behavior introduces yet another important problem: if there are several clients attempting to connect to a given destination on the same port 
(e.g., $\left\{ {IP_{B}, port_{B}} \right\}$) through switch $R$, the latter will need to use a different port for each outgoing connection.
However, since TCP port numbers are 16~bit fields, the maximum number of connections the switch will be able to migrate to a given IP-port pair 
is at most $2^{16} - 1024 = 64512$ (the first 1024 ports are reserved for well known services).
This number can be quickly reached if we consider extremely popular HTTP services (e.g., Google or Facebook). 
Therefore, each switch is bound to a maximum number of connections it can migrate for each service, 
after which all new incoming connection requests can not be satisfied.

The limited number of available ports can easily be exploited by an attacker to mount DoS attacks. 
In fact, it would be easy to target a given service by opening a series of long-lasting connections through the OF switch. 
Once enough connections are opened and all the possible ports have been used by the switch to migrate the connections, 
any other client trying to connect to said service will be rejected.
This bound on the maximum number of connections that a switch can migrate, provides a simple and efficient 
way of mounting a DoS attack to a given host $B$, for all the clients whose path to $B$ passes through the same OF switch.
There are no definitive solutions to this problem if proxying is used: each connection to the same service must be assigned a unique port number by the switch.
The most promising way to somewhat mitigate this restriction is to purchase several IP addresses for the switch to use. 
In this way, when the OF switch consumes all the available ports with a given address, it will switch to a new address, this way being able to migrate other $64512$ connections.
While address purchasing can be employed as a partial solution, it is worth noting that for each additional address, the space of possible combinations increases by just $64512$. 
If we consider a complex network, where several switches employ the connection migration technique, 
we will quickly hit a point where the cost and complexity of management increase extremely rapidly, reducing the appeal of this workaround.


\section{Our Solution: LineSwitch}\label{sec:lineswitch}

Breaking TCP end-to-end semantics introduces the need to store state, which in turn opens the system to attacks exploiting buffer saturation. 
Therefore, there is a strong need to reduce its use as much as possible while retaining its beneficial effects against control plane saturation attacks.
To reach this goal, we propose LineSwitch, an OpenFlow extension intended for edge OF switches of a network. 
The idea is the following: LineSwitch proxies all incoming TCP connections from a given IP until one is completed, while for all subsequent SYN packets the proxy is used with a small probability $P_p$.

Our solution effectively protects the switch even in presence of an attacker $E$ with knowledge about the proxy mechanism in use.
Indeed, it would be possible for $E$ to perform an attack by correctly completing the handshake associated with the first SYN packet sent, and then initiating the SYN flooding.
Although this is true, $E$ would be forced to use its real IP address in all the packets, to ensure they will be forwarded to the OF pipeline. 
Otherwise, they will be proxied by the switch, and therefore discarded.
Moreover, since the effectiveness of the SYN flooding attack is based on a high throughput, 
once $E$ is forced to use its real IP address for the flooding, 
the OF switch will quickly proxy one of the packets, thus detecting the attack.
Then, the OF switch can blacklist the IP address of host $E$, $IP_E$, for $T\times2^{count_{IP_E}}$~seconds, where $count_{IP_E}$ indicates the number of times $IP_E$ did not complete a connection, and $T$ represents a default time value.
	
With our approach, all packets with spoofed IP addresses will be blocked at the data plane, 
and malicious clients that initiate SYN flooding with non-spoofed IPs (after first establishing a complete TCP connection) 
will be penalized with rapidly increasing blacklist periods of time. 

Our approach provides two major improvements with respect to the problems identified 
in the state-of-the-art (see Section~\ref{sec:connection_migration_analysis}).
The first advantage of our approach is that it drastically reduces the memory usage needed to perform the address/port translation for each connection, 
thus offering protection against the described buffer saturation attack (see Section~\ref{sec:proxying_needs_state}).
Indeed, LineSwitch requires port translation only for the first SYN packet per each IP address, and only for a very small number of packets after that based on the chosen probability $P_p$. Therefore, the memory usage increases roughly linearly with the number of clients with a TCP connection through the switch R. In contrast, the memory overhead introduced by \avantguard~grows linearly {\em with the number of connections} that passe through the switch.
An attacker can easily generate a huge number of connections from the same source IP using different port numbers (up to $2^{16}=65535$ per $<IP_{dst}, port_{dst}>$ pair; the theoretical limit would then be $2^{16+32+16}$) and, with \avantguard, the switch would need to store state for each of these connections. 
As a reference, in our experiments with a link of 1~Mbps we were able to open approximately 780 connections per second, while with a higher bandwidth of 5~Mbps, it was possible to complete more than 4000 connections per second.
Instead, with LineSwitch the number of entries the switch needs to store under attack, is roughly proportional to the number of distinct real IP addresses (machines) the attacker possesses. As a consequence, the effect of buffer saturation attacks is greatly reduced, while at the same time retaining full protection against SYN flooding attacks.

A second major advantage with respect to the state-of-the-art is that LineSwitch proxies a minimum number of connections.
Indeed, as discussed in Section~\ref{sec:connection_migration_analysis}, while being an effective mechanism to protect against SYN flooding attacks, 
proxying introduces several problems which derive from breaking the end-to-end paradigm.
Therefore, its use should be limited as much as possible. LineSwitch proxies only the first connection from a given host (i.e., an IP address), and subsequent connections from the same host with probability $P_p$. 
Since, in general, the migration probability value $P_p$ is small (see Section~\ref{sec:evaluation}), LineSwitch will assure in most cases the normal network flow, mitigating intrinsic problems of proxying such as limited maximum number of migrated connections (see Section~\ref{sec:proxy_ports}).
 Moreover, our experimental results show that a small value for $P_p$ does not reduce the level of protection that LineSwitch provides against the control plane saturation attack (see Section~\ref{sec:evaluation}).


\section{Preliminary Evaluation}\label{sec:evaluation}

\begin{figure}[h!t]
	    \centering
	    \includegraphics[width=0.6\linewidth]{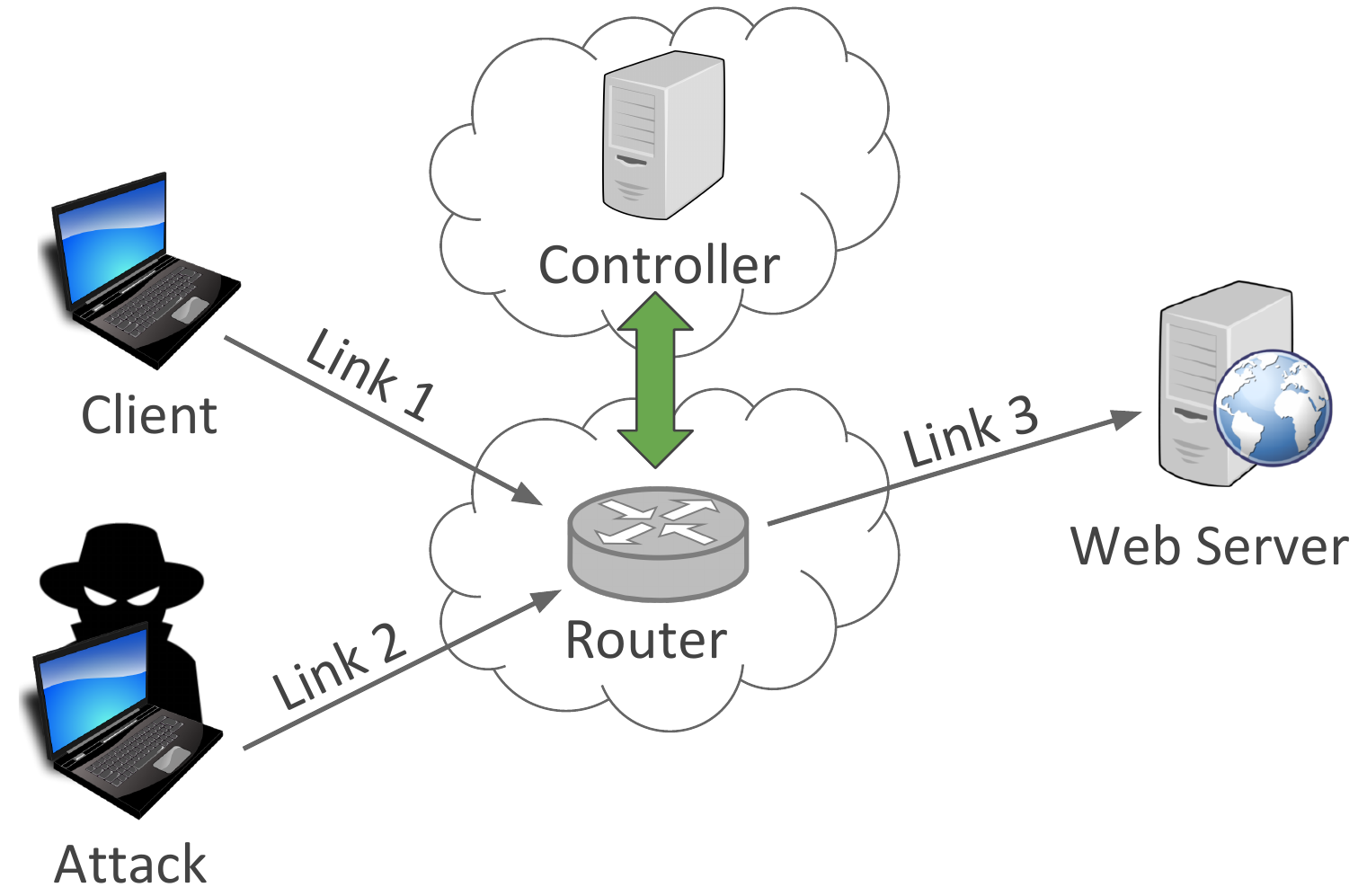}
		\caption{Experimental Setup.}
	    \label{fig:sim_setup}
	\end{figure}

In order to assess the effectiveness of LineSwitch against the control plane saturation attack, and its resiliency to buffer saturation attacks, we designed preliminary experiments on the setting shown in Figure~\ref{fig:sim_setup}. Our system model includes three hosts connected to an OF switch running the reference OpenFlow software switch~\cite{RefOFSwitch}, and a local controller (running the POX controller, {\tt l3\_learning} module~\cite{POX}).
We ran all our experiments using the Mininet network simulator~\cite{Mininet} in a virtual machine.
We compared LineSwitch to the state-of-the-art solution to tackle SYN flooding-based control plane saturation attack, i.e., \avantguard~\cite{AvantGuard}.
The computer used for the simulation is equipped with a quad core Intel i5-4670 @3.40GHz, all of which were available to the virtual machine.
	
We simulated the behavior of both \avantguard~and LineSwitch under the buffer saturation attack introduced in Section~\ref{sec:lineswitch}.
To this aim, we configured the system with different buffer sizes and run the attack at different rates. 
As a result, we demonstrate that: (1) the attack rate required to successfully incapacitate an OpenFlow switch running \avantguard~grows {\em linearly with the size of the buffer}; (2) when using \avantguard, the throughput needed to successfully complete the attack in a reasonable amount of time {\em is easily achievable, even with larger buffers}; (3) LineSwitch offers an extremely high resiliency to the buffer saturation attack, and can be further configured through the $P_p$ parameter to address the specific needs of the network.

\begin{figure}[h!t]
	    \centering
		\begin{subfigure}{0.49\columnwidth}
			\centering
			\includegraphics[width=\columnwidth]{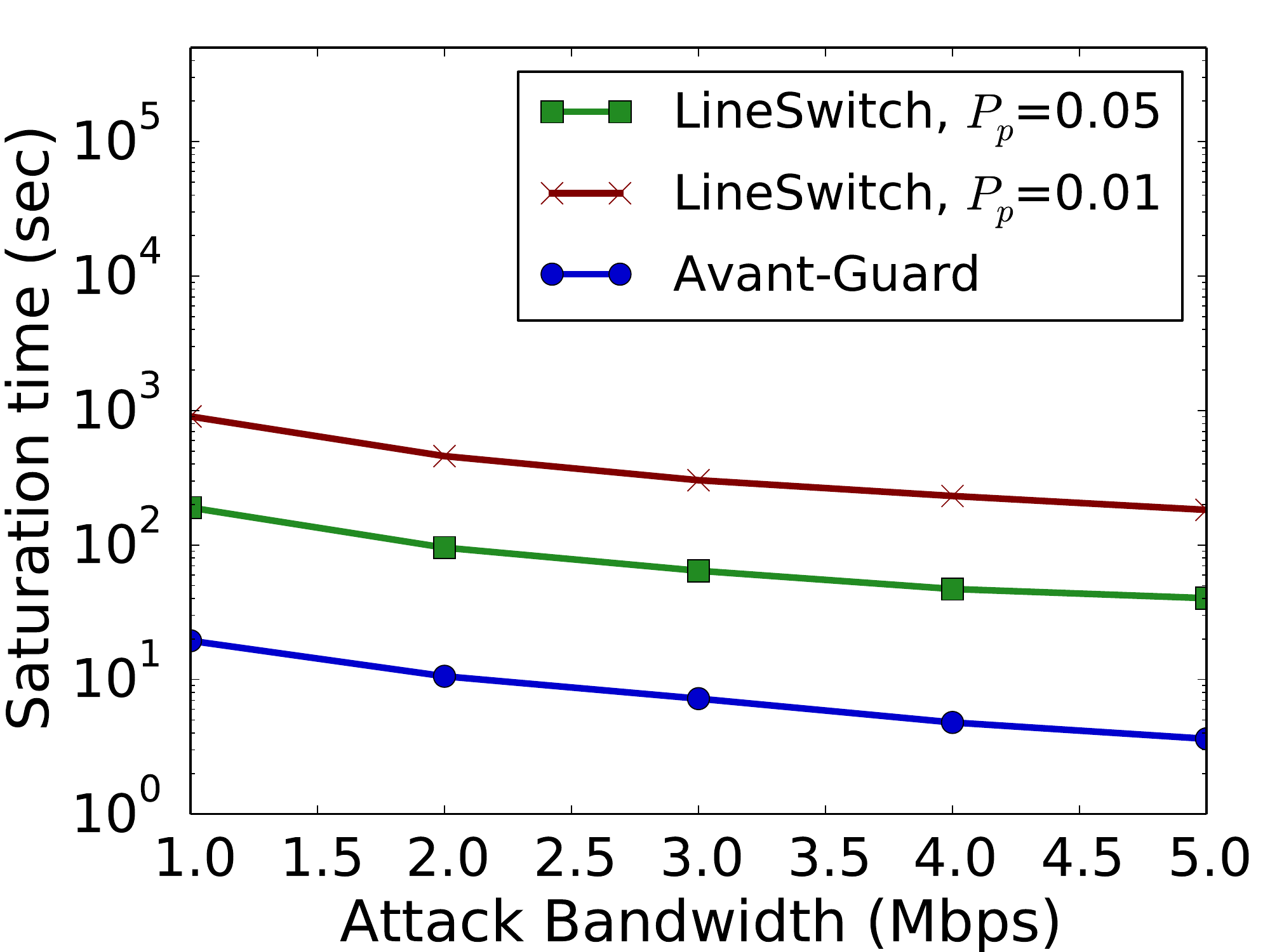}
			\caption{buffer size $2^{20}$Bytes}
			\label{fig:graph20}
		\end{subfigure}
		\begin{subfigure}{0.49\columnwidth}
			\centering
			\includegraphics[width=\columnwidth]{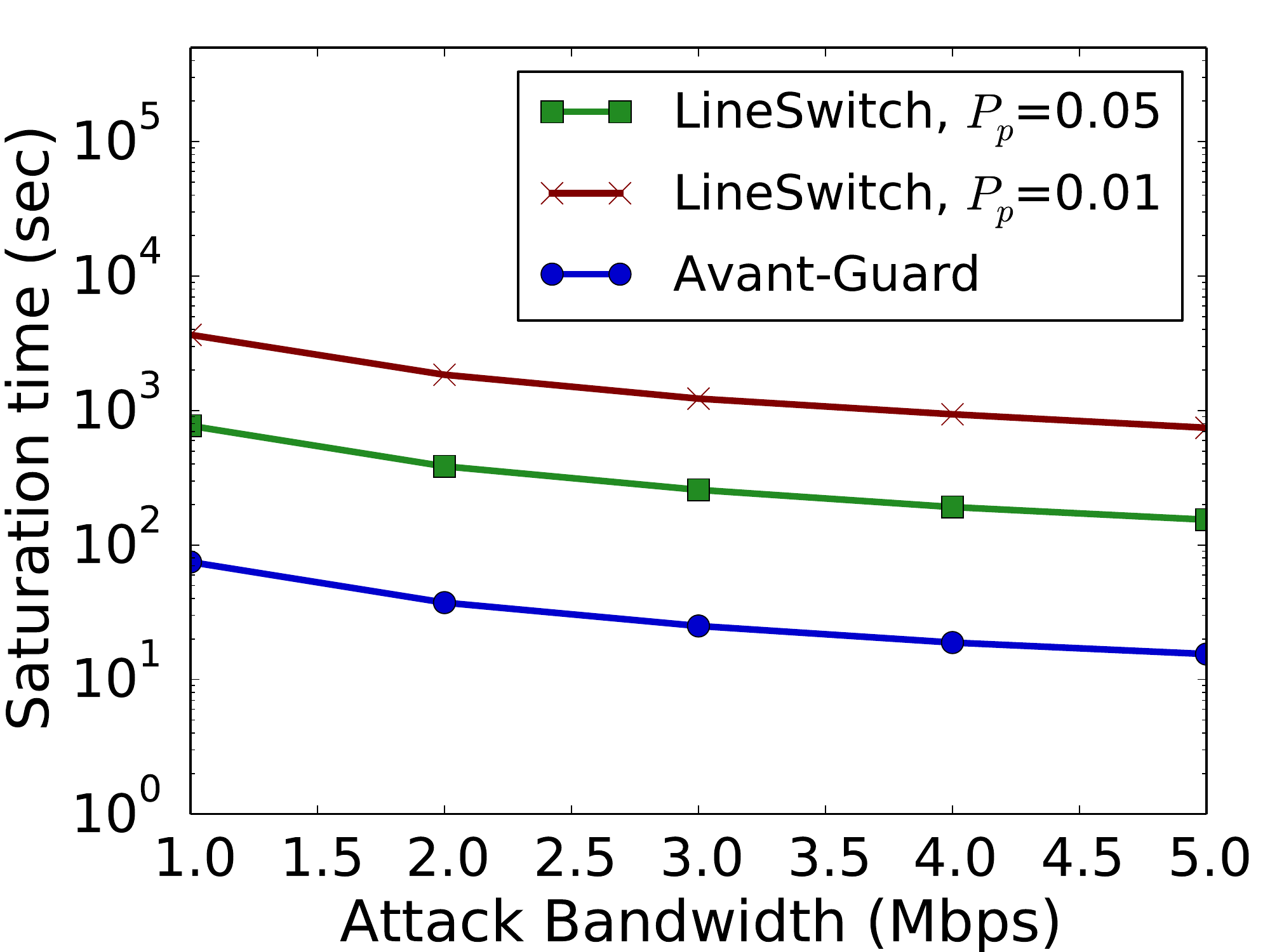}
			\caption{buffer size $2^{22}$Bytes}
			\label{fig:graph22}
		\end{subfigure}
		\caption{Average buffer saturation time, varying buffer size and attack bandwidth, under a buffer saturation attack. Time on y-axis is in logarithmic scale.}\label{fig:time_saturation}
	\end{figure}

In order to directly assess the relation between attack bandwidth and time required to saturate the buffer, we set the RTT is set to 0ms, for each link in Figure~\ref{fig:sim_setup}.
In any case, since the attacker continuously floods the switch with new connections, the RTT would have been relevant just until the first connection were completed.
Figure~\ref{fig:time_saturation} presents the results of our simulation. It shows the average time needed to successfully overload a switch with a buffer saturation attack, running both \avantguard~and LineSwitch, with the latter executed with parameter $P_p$ set to $0.01$ and $0.05$. 
The results are presented for varying size of the buffer (expressed in Bytes) and for different rates of attack (expressed in Mbps).

As Figure~\ref{fig:time_saturation} shows, even with a modest rate of attack it is possible to quickly overflow the buffer of a switch running \avantguard: with an attack rate of 1~Mbps, a buffer of $2^{22}$~Bytes is saturated in 74.718~sec, preventing the switch from migrating any new connection.
By contrast when using LineSwitch, even when setup with a highly conservative migration probability $P_p=0.05$, the time needed to perform a successful buffer saturation attack is one order of magnitude greater when compared to \avantguard. 
As an example, with a 1~Mbps attack rate, a buffer size of $2^{22}$~Bytes and $P_p=0.05$, LineSwitch requires 769.487~sec to be saturated against only 74.718~sec required when running \avantguard. 
When using lower (and more realistic) migration probability values, the time difference increases even more, as shown in Figure~\ref{fig:time_saturation}.
 
For completeness, we compared the average overhead introduced by \avantguard~with the overhead introduced by LineSwitch, evaluating both a scenario without attack, and a scenario under SYN flooding based control plane saturation attack.
All overhead data is expressed with respect to the standard OpenFlow protocol, under normal network conditions (e.g., no attacks performed).
We sampled the time required by the legitimate client (see Figure~\ref{fig:sim_setup}) to download a web page of size 1~KB from the web server. 
In the regular traffic scenario, \avantguard~introduces an average time overhead of 41.83\%, while LineSwitch incurs only a 7.67\% overhead. Moreover, under control plane saturation attack with an attack rate of 6.5~Mbps, \avantguard~introduces an overhead of 36.92\%, while LineSwitch introduces only a 5.45\% overhead. 
Finally, under control plane saturation attack, both \avantguard~and LineSwitch guarantee a 100\% page retrieval success rate.


\section{Conclusion}\label{sec:conclusion}
In this paper we analyzed the effects of the control plane saturation attack based on SYN flooding, one of the most widespread types of denial of service attack, 
when applied to Software Defined Networks (SDN) architecture, and in particular to its reference implementation, OpenFlow.
We showed that the extensive communication needed by the control plane and the data plane in SDN amplifies the effect of typical denial of service attacks, resulting in an overload of the control plane and in the possible impairment of large parts of the network. 
Furthermore we considered \avantguard~\cite{AvantGuard}, which is, to the best of our knowledge, the only currently proposed solution against control plane saturation attack. We showed that in its original design, subtle points were not taken into consideration, opening critical system vulnerabilities.
To this aim we proposed LineSwitch, a solution based on probability and blacklisting which offers both resiliency against SYN flooding-based control plane saturation attacks and protection from buffer saturation vulnerabilities.
Our preliminary evaluation demonstrates that LineSwitch imposes a negligible overhead, which can be dynamically adjusted to fit the network needs, 
while successfully defending the OpenFlow switch and controller from attacks that can potentially disrupt the functionality of the network.

\balance
\bibliographystyle{abbrv}
\bibliography{bibliography}

\end{document}